\documentclass{article}

\usepackage[english]{babel}
\usepackage[letterpaper,top=2cm,bottom=2cm,left=3cm,right=3cm,marginparwidth=1.75cm]{geometry}
\usepackage{amsmath}
\usepackage{graphicx}
\usepackage[ruled]{algorithm2e}
\usepackage{multirow}
\usepackage{dingbat} 
\usepackage[pagebackref=true,breaklinks,colorlinks=true, allcolors=blue]{hyperref}
\usepackage{adjustbox}
\usepackage{authblk}
\usepackage[justification=centering]{caption}

\title{An Evaluation and Comparison of GPU Hardware and Solver Libraries for Accelerating the OPM Flow Reservoir Simulator}

\author[1]{Tong Dong Qiu}
\author[2]{Andreas Thune}
\author[3]{Vinícius Oliveira Martins}
\author[4]{Markus Blatt}
\author[5]{\\Alf Birger Rustad}
\author[1]{R\u{a}zvan Nane}
\affil[1]{ Big Data Accelerate B.V., The Netherlands}
\affil[2]{Simula Research Laboratory, Norway}
\affil[3]{Universidade Estadual de Campinas, Brazil}
\affil[4]{OPM-OS AS, Norway}
\affil[5]{Equinor S.A., Norway}

\begin{document}
\maketitle

\begin{abstract}
Realistic reservoir simulation is known to be prohibitively expensive in terms of computation time when increasing the accuracy of the simulation or enlarging the model grid size. One method to address this issue is to parallelize the computation by dividing the model into several partitions and using multiple CPUs to compute the result using techniques such as MPI and multi-threading. Alternatively, GPUs are also a good candidate for accelerating computation due to their massively parallel architecture, which allows many floating point operations per second to be performed. Although Computational Flow Dynamics problems are complex and contain many computational parts that are challenging, the most difficult one is the execution of the numerical iterative solver that is used to solve the linear systems, which arise after discretizing the nonlinear system of equations that mathematically model the problem. The numerical iterative solver takes thus the most computational time and is challenging to solve efficiently due to the dependencies that exist in the model between cells. In this work, we evaluate the OPM Flow simulator and compare several state-of-the-art GPU solver libraries as well as custom-developed solutions for a BiCGStab solver using an ILU0 preconditioner and benchmark their performance against the default DUNE library implementation running on multiple CPU processors using MPI. The evaluated GPU software libraries include a manual linear solver in OpenCL and the integration of several third-party sparse linear algebra libraries, such as cuSparse, rocSparse, and amgcl. To perform our bench-marking, we use small, medium, and large use cases, starting with the public test case NORNE that includes approximately 50k active cells, and ending with a large model that includes approximately 1 million active cells. We find that a GPU can accelerate a single dual-threaded MPI process up to 5.6 times and that it can compare with around 8 dual-threaded MPI processes.

\end{abstract}

\section{Introduction}

Computational Flow Dynamics (CFD) simulation is becoming extremely important to speed up prototyping and validating new designs for large-scale models, which are difficult to develop by a purely analytical method. For instance, reservoir simulators are used extensively by reservoir engineers nowadays to analyze the flow of fluids and predict oil and gas production. This applies both to managing current and developing new fields. The core of such a CFD simulation is finding the solution to a set of partial differential equations (PDEs) constrained on some domain of interest. This is a two-step process, where the PDEs are first discretized over the defined domain using numerical techniques such as the finite-difference, -element or -volume method, and second, solving the obtained linear and nonlinear systems using iterative linear solvers such as Krylov subspace solvers. However, solving the linear and nonlinear systems produced by these finite methods is usually time-consuming and complex because the linear systems are sparse and ill-conditioned. Furthermore, to improve the realism of the simulation, the number of cells required to be modeled is increasing at a steady pace. This leads to bigger systems, which in turn leads to higher run times. To accommodate these bigger systems and improve the scalability of the simulation, parallelization can be performed in two major directions: multi-core CPUs and many-core GPUs. In this work, we focus on the latter using the OPM project.

The Open Porous Media (OPM) \cite{OPM} project encourages open and reproducible research for modeling and simulation of porous media. It was chosen as the base code for our work here since it is the only open implementation of a reservoir simulator capable of running industrially relevant reservoir simulations. OPM supports several types of reservoir simulation including black oil, which is the most commonly used fluid model in the energy industry. The project interfaces state-of-the-art scientific libraries such as the DUNE \cite{dune} to perform the computations efficiently. It leverages powerful Krylov subspace preconditioners and solvers that are generally used to solve large-scale and sparse linear systems. Preconditioning and linear solutions typically comprise 50-90\% of total simulation time depending on the case and methods used. Hence, this part is our prime target for acceleration on GPU. At the time of this work, the best-performing linear solver on the CPU was BiCGSTAB from the DUNE library. For the preconditioner, the ILU0 performed best in OPM Flow. This was demonstrated in \cite{ILU0-opt}, where OPM Flow was benchmarked against the industry leading simulators Eclipse 100 and Intersect on the openly available Norne full field model. BiCGSTAB and ILU0 outperformed CPR ( contrained pressure residual, see \cite{CPR}) with AMG (algebraic multigrid) both on single process performance as well as scaling. Hence, the focus of our work here is to investigate how a GPU implementation of the linear solver and preconditioner will perform. 

GPUs have many small cores that can work together using the Same Instruction Multiple Data (SIMD) model. Each core performs the same instruction, on different data. This allows the GPU to process massive amounts of data, much faster than a CPU would, but only if the problem has enough parallelism. GPUs have shown large speedups in many different fields, like weather simulation\cite{gpu_weather}, FEM-based structural analysis\cite{gpu_struct}, numerically integrating ordinary differential equations (ODEs)\cite{gpu_ode}, chip design\cite{gpu_chip} or machine learning with Tensorflow\cite{gpu_tensorflow}. Although GPUs have been highly effective in many application domains, their usefulness in reservoir simulation remains uncertain due to specific sparsity characteristics. In particular, the different sparsity patterns and the dependencies between reservoir model cells cause code divergence. As a result, cells placed into independent partitions, due to dependencies, are still executed sequentially. To analyze how big this problem is, we develop, integrate, and perform a thorough evaluation with both manual OpenCL kernels and various third-party libraries targeted at different GPU hardware to accelerate the linear solver of OPM Flow\cite{flow_paper} on a GPU. We target the default OPM ILU0 preconditioned BiCGStab linear solver used in OPM Flow. Concretely, the novelty of the paper is summarized as follows:

\begin{itemize}
    \item Develop open-source manual OpenCL and CUDA kernels for ILU0 preconditioned BICGStab solver.
    \item Develop a custom open-source bridge to integrate the custom developed solvers as well as third-party libraries into the OPM Flow simulator. 
    \item Integrate both manual solvers and third-party libraries into OPM Flow, including amgcl, cuSparse, and rocSparse.
    \item Perform a thorough evaluation and comparison of all the developed and integrated libraries using three reservoir models with different sizes and running on GPU hardware from both Nvidia and AMD vendors. This is, to the best of our knowledge, the first work to present complete GPU results running on the open-source reservoir simulator using real-world use cases. 
\end{itemize}

The paper is organized as follows. First, in section \ref{sec:background} we describe the background required to understand the reservoir simulator source code and the GPU preconditioners and iterative solver developed in this work. Then, in section \ref{sec:relres} we highlight other interesting reservoir simulators and indicate what target processors they support. Section \ref{sec:implementation} describes the implementation details of how the different manual and third-party libraries were developed and integrated, including particularities specific to reservoir simulation such as well contributions to the final solution. Section \ref{sec:experimentalresults} presents the experimental results and discusses the obtained performance numbers for the different libraries and GPU hardware tested. Finally, section \ref{sec:conclusion} summarizes the paper and lists future work. 

\section{Background}
\label{sec:background}

In this section, we will describe the OPM project and introduce the basic concepts used in the CFD field of reservoir simulation, such as grid, assembly, linear iteration, preconditioner, and linear solver. Also, concepts specific to reservoir simulation such as the reservoir wells are defined.

\subsection{The open source code base}

Reservoir simulation\cite{reservoir_sim1}\cite{reservoir_sim2} uses mathematical models to predict the fluid flow dynamics in porous media. Typically, oil, water, and gas are modeled as fluids and the porous media are usually rock or soil. The simulations can be used to improve reserve estimates, predict future production, or evaluate multiple reservoir management strategies. The first step is to create a static model of the reservoir, also called a geological model. Geologists and geophysicists create this geological model by using multiple types of information, such as seismic data, well logs, production history, and rock properties. Important rock properties are type, porosity, water saturation, and permeability. Recall, that we chose to use OPM as a starting point for our implementations because it is the only open source reservoir simulator capable of running realistic reservoir models. However, all implementations are open, so all results presented are general and can be ported to or re-implemented in any code base (provided the software license is not violated). Hence, we will give an overview of the reservoir simulator implementation, but all capable reservoir simulators share a similar structure. This will hopefully enable peers to easily reproduce and further improve our results here.

The Open Porous Media (OPM) \cite{OPM} simulator, OPM Flow \cite{flow_paper}, is an open-source simulator that models black-oil \cite{blackoil_model} with dissolved gas and vaporized oil. Other characteristics of the model that can be modeled include rock-dependent capillary and relative-permeability curves, end-point scaling and hysteresis, and oil vaporization controls. The simulator's input file is compatible with the commercial simulator Eclipse \cite{eclipse}, and the output file is readable by commercial post-processing software. The OPM project also features a post-processing software called ResInsight \cite{resinsight}. The black-oil model assumes three fluid phases (aqueous, oleic, and gaseous) and three components: water, oil, and gas. The oil and gas components represent all hydrocarbons in liquid and vapor form at standard conditions respectively. Mixing is possible, so both oil and gas can be found in the oleic phase, and gaseous phase of both. The partial differential equations (PDEs) are derived from the conservation of mass and Darcy's law\cite{darcy_law1}\cite{darcy_law2}, together with suitable initial and boundary conditions. The equations give each grid element three unknowns. Additionally, the wells each have equations and unknowns as well. Choosing which unknowns to solve for is important. For non-miscible flow, the oil pressure $p_o$, water saturation $s_w$, and gas saturation $s_g$ are chosen. For miscible flow, the gaseous phase may disappear if all the gas dissolves into the oleic phase. The oleic phase can also disappear if all the oil vaporizes into the gaseous phase. The oil pressure and water saturation are chosen, but the third variable is flexible:

\begin{equation*}
    x = \begin{cases}
        s_g,      & \text{all three phases present,} \\
       r_{go},   & \text{no gaseous phase,} \\
      r_{og},  & \text{no oleic phase,} \\
    \end{cases}
\end{equation*}

with $r_{go}$ being the ratio of dissolved gas to oil in the oleic phase, also called $r_S$ in other literature, $r_{og}$ being the ratio of vaporized oil to gas in the gaseous phase, also called $r_V$ in other literature.

\begin{figure}[h]
\centering
\includegraphics[scale=0.4]{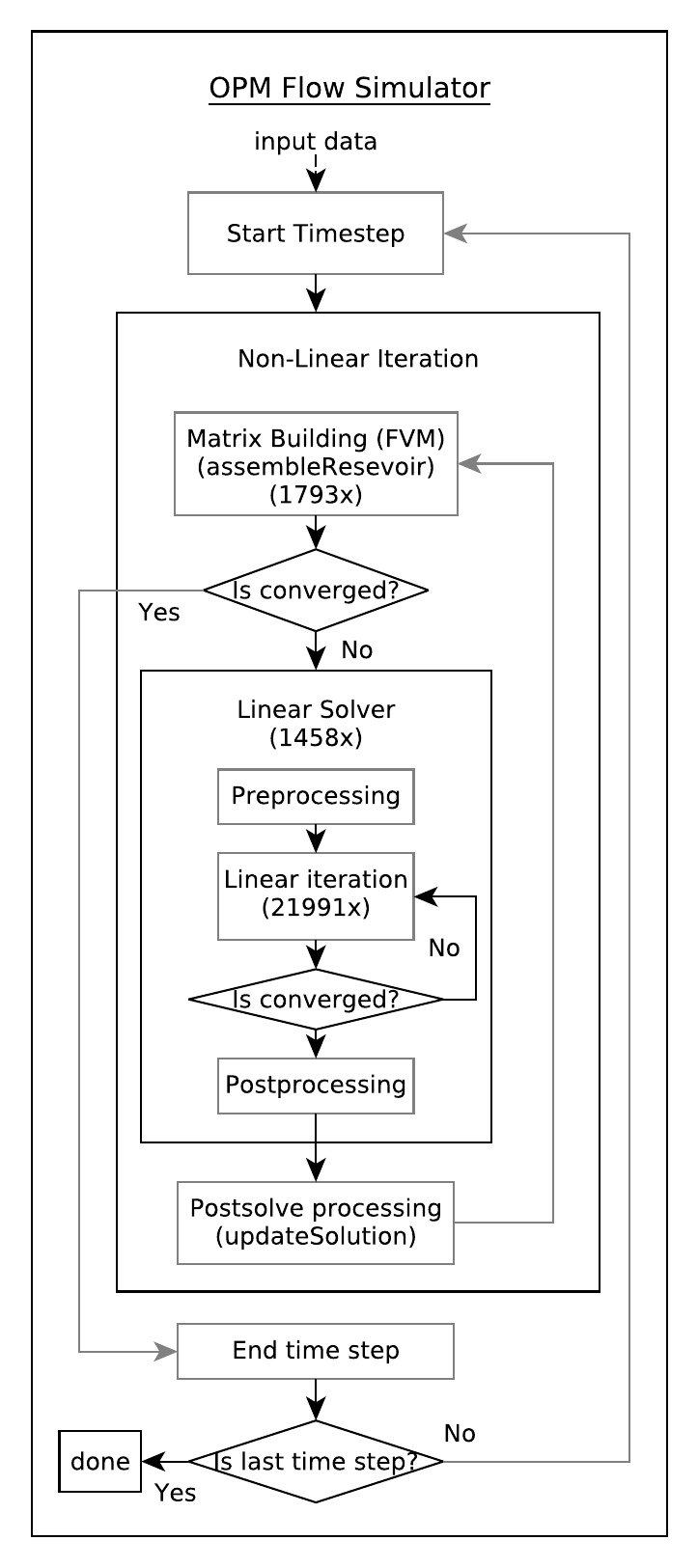}
\caption{The General Structure of OPM Flow.}
\label{fig:flow_chart}
\end{figure}

The PDEs need to be discretized to be solved numerically. They are discretized in space with an upwind finite-volume scheme, with a two-point flux approximation.
Discretization in time is done using an implicit backward Euler scheme. The derived equations form a system of fully implicit nonlinear equations. This system is solved using a Newton-Raphson method and linearized. The resulting linear system is solved with a preconditioner linear solver. Figure \ref{fig:flow_chart} shows the general structure of OPM Flow.

The reservoir is modeled by a grid, where each cell/element has its own properties, such as porosity, volume, or transmissibility. The grid properties are defined in input files, which include structure, faults, and various static rock properties like porosity and permeability. OPM supports 1D, 2D, and 3D models, with three types of grids: Cartesian Regular, Radial, and Irregular Corner-Point. The first two types are relatively simple but are also rather limited. Irregular Corner-Point grids are the industry standard for describing the structure of complex reservoirs. The Cartesian Regular grid defines a regular orthogonal grid. For Irregular Corner-Point grids, coordinates lines or pillars are given to indicate x and y coordinates. Then the top and bottom surfaces are specified by the z-coordinates of the cell's corner points along the four adjacent pillars. The cell then forms an irregular hexahedron, with each cell having the same outline as the cell above or below when viewed from above. Radial grids can be used to model radial flow near a wellbore. The Radial grids in OPM are currently modeled as 2D cylindrical grids because no flow in the theta direction is implemented. Grids cannot be combined in OPM, so Irregular Corner-Point is mostly used. For more details on these grids, see the OPM Flow Reference Manual\cite{flow_manual}.

The simulation is fully implicit in time and has a flexible assembly of the linear system through automatic differentiation to enable the rapid development of new fluid models. Traditionally, the closed-form expressions are obtained by differentiating the discretized flow equations by hand. This process is time-consuming and error-prone, but by using automatic differentiation these drawbacks are removed. The simulator provides adaptive time step size controls. OPM Flow only uses Finite Volume Method (FVM), but the OPM project as a whole also implements Finite Element Method (FEM).
 The assembled linear system is blocked and sparse. The matrix is stored in the BCRSMatrix (block compressed row storage) data structure, provided by dune-istl \cite{dune-istl}. Each block is a FieldMatrix, a dense matrix provided by dune-common. Due to fluid properties and other reservoir-specific parameters, the discretized equations could show elliptic, parabolic or even hyperbolic behavior. The resulting linear system is usually non-symmetric and ill-conditioned.
Instead of solving the system directly, the solution is approximated by a preconditioned iterative solver. The solver can either be bicgstab (default) or restarted GMRES. For the preconditioner, the options are ILU0 (default), AMG, or CPR.

OPM features two different well models: \textit{standard} and \textit{multi-segment}. A standard well has a single set of primary variables to describe the flow conditions inside. This works adequately for most wells and is therefore the default well model. For a three-phase black oil system, there are four primary variables: the weighted total flow rate, the weighted fractions of water and gas, and the bottom-hole pressure. 
 Each well object has three variables in the implementation that are used during the linear solve: B, C, and D. B and C are 1xNb blocked, sparse matrices, with MxN blocks, where Nb is the number of blockrows of the matrix, and M and N are 4 and 3 for blackoil respectively. D is a small MxM matrix. To apply a standard well, the solution vector x is a $C^T * (D^{-1} * (B * x))$. The multi-segment wells are used to simulate more advanced wells, such as multilateral wells, horizontal wells, and inflow control devices. The wellbore is divided into a number of segments, where each segment consists of a segment node and a flow path to the neighboring segment in the direction of the well head (outlet segment). Most segments also have inlet segment neighbors. Each segment has, in addition to the primary variables of a standard well, a node pressure variable.

Finally, table \ref{tab:OPM_modules} lists the different OPM modules and their description.
\begin{table}[h!]
\centering
\begin{tabular}{c|p{12.5cm}}
Name & Description \\ \hline
opm-common     & reads Eclipse data files, provides Python bindings, builds systems \\
opm-grid       & provides an interface for Dune-grid \\
opm-material   & deprecated, now inside opm-common, provided infrastructure to handle material properties like relative-permeability/capillary pressure, thermodynamic relations, and empirical heat conduction laws \\
opm-models     & contains fully-implicit numerical models for flow and transport in porous media \\
opm-simulators & contains simulator programs, like Flow, a fully implicit black-oil simulator that supports solvent and polymer options \\
\end{tabular}
\caption{Different OPM Modules and their Description.}
\label{tab:OPM_modules}
\end{table}

\subsection{ILU0}
To solve the linear system $Ax=b$ efficiently, the matrix $A$ is approximated by a factorization $A\approx LU$, with lower unitriangluar matrix $L$ and upper triangular matrix $U$. Once we obtained such a LU factorization, solving the linear system is performed according to the equations (3) and (4), which is equivalent to solving (1), both methods leading to finding the unknown $x$ vector:

\begin{align}
    Ax &= b \\
    LUx &= b \\
    Ly &= b \\
    Ux &= y
\end{align}

First equation (3) is solved using a forward substitution, and second, with the solution of $y$, a backward substitution is performed as in equation (4) to find $x$. The linear solves $Ly=b$ and $Ux=y$ are relatively easy since L and U are triangular matrices. In OPM terminology, this is called ILU0 application, and algorithm \ref{alg:ilu_apply} highlights this process.


\begin{algorithm}
\caption{ILU0 application}
\label{alg:ilu_apply}
Input: vector x, vector b, matrix L, matrix U \\
Output: updated vector x \\

// forward substitution

// backward substitution

\end{algorithm}

For ILU0, L and U have the same sparsity pattern as matrix A. To find the $L$ and $U$ matrixes, a simple and sequential method to perform the decomposition is used that is listed in Algorithm \ref{alg:ilu_decomp}.

\begin{algorithm}
\caption{ILU0 decomposition}
\label{alg:ilu_decomp}
Input: matrix A \\
Output: matrix A, but contains both L and U \\

// nb is the number of blockrows of A \\
for r = 0 to nb: \\
    d = 1/$a_{rr}$ \\
    for i = r+1 to nb: \\
        if (i,r) exists in S: \\
            e = d*$a_{ir}$ \\
            $a_{ir}$ = e \\
            for j = r+1 to nb: \\
                if (i,j) and (r,j) exist in S: \\
                    $a_{ij}$ = $a_{ij}$ - e*$a_{rj}$ \\
\end{algorithm}

\subsection {GPU Architecture and Programming Model}
GPUs are traditionally designed for display purposes.
Since each pixel is independent of others, they can be processed in parallel.
This means the GPU has a parallel architecture.
It can also process other algorithms that have parallelism.

GPUs have a massively parallel architecture, which allows them to process large amounts of data in a relatively short time if the processing algorithm has enough parallelism.
The computation is done by stream processing.
Data can be read, processed and written at the same time, in a pipeline.
The two big manufacturers are NVIDIA and AMD.
NVIDIA proprietary language CUDA is only available for NVIDIA GPUs, and the open-source OpenCL can run on both.
For a kernel, a set of threads are launched, called workitems.
Workitems are organized in workgroups.
Workgroups can operate independent from each other, but the workitems in a workgroup are mapped to the same hardware area, a Streaming Multiprocessor (SM) on NVIDIA GPUs, or a Compute Unit (CU) on AMD GPUs.
The workitems are also divided in wavefronts of 64 workitems.
All 64 workitems execute the same instructions at the same time.
If the kernel contains a branch that is not taken by all 64 workitems, the branch is serialized.
Some of the workitems are deactivated, while the other take their branch, then activation is swapped and the other branch is taken.
For NVIDIA GPUs, the wavefronts consist of 32 workitems.
Table \ref{tab:cuda_opencl_glossary} defines some of the OpenCL programming concepts and their CUDA counterparts.

It is important to only perform coalesced global memory accesses.
The workitems all perform the same global memory read/write instruction, but with slightly different addresses.
If the addresses are coalesced, the memory reads are combined into as many separate, serialized transactions as are needed.
If all workitems read from the same memory chunk, it only needs one transaction.

\begin{table}[h]
\centering
\begin{tabular}{c|c}
OpenCL & CUDA \\ \hline
wavefront & warp \\
workgroup & (thread)block \\
workitem & thread \\
local memory & shared memory \\
private memory & local memory \\
global memory & global memory \\
Compute Unit (CU) & Streaming Multiprocessor (SM)
\end{tabular}
\caption{Some OpenCL defintions and their CUDA counterparts.}
\label{tab:cuda_opencl_glossary}
\end{table}

\section{Related works}
\label{sec:relres}

Other reservoir simulators include:
\begin{itemize}
\item BOAST\cite{boast1}\cite{boast2}: Black Oil Applied Simulation Tool is a free simulator from the U.S. Department of Energy. It uses IMPES (finite difference, implicit pressure, explicit saturation). The last release was in 1998.
\item MRST\cite{mrst}: Matlab Reservoir Simulation Toolbox is developed by SINTEF Applied Mathematics, who also contributed to OPM. It also includes third-party modules from developers from many different research institutes.
\item ECLIPSE\cite{eclipse}: Originally developed by ECL (Exploration Consultants Limited), now owned and developed by Schlumberger, Eclipse is an industry reference simulator. The in- and output files of OPM are compatible with Eclipse software.
\item INTERSECT\cite{intersect}: Also from Schlumberger, it has similar features as Eclipse. Intersect has more support for massively parallel execution, with a GPU-accelerated linear solver, and faster linearization of equations.
\item ECHELON\cite{echelon}: Echelon is the only reservoir simulator that is fully GPU accelerated. It uses a GMRES solver with a CPR preconditioner. A maximum of 12 million active cells can be simulated on a single GPU.
\item TeraPOWERS\cite{terapowers}: TeraPOWERS is an in-house simulator by Aramco. It boasts the world's first trillion cell simulation, performed on 150000 cores of the Shaheen II supercomputer\cite{shaheen}.
\item GEOSX\cite{geosx}: An open-source simulator, specifically for modeling carbon storage. It does not appear to have a three-phase black oil simulation. There is a possibility of using CUDA, but it is unclear which parts are accelerated exactly.
\item tNavigator\cite{tnavigator}: A black oil, compositional, and thermal reservoir simulator. Allows the linear solver part to run on a GPU, as well as some other parts.
\item PFLOTRAN\cite{pflotran}: An open source simulator that uses PETSc\cite{petsc} for domain decomposition to achieve parallelism. Not accelerated on GPU, but is able to scale easily on CPUs.
\end{itemize}

Numerous libraries that support blocked sparse linear algebra on GPU exist, including cusparse\cite{cusparse}, magma\cite{magma}\cite{magma_paper}, viennacl\cite{viennacl}, ginkgo\cite{ginkgo-toms-2022}, amgcl\cite{amgcl1}, rocalution\cite{rocalution} and rocsparse\cite{rocsparse}.
Of these, only cusparse does not support AMG GPUs at all, others might need the HIP\cite{hip} module from the ROCm\cite{rocm} framework.
Rocalution actually uses many functions from rocsparse underneath.
Table \ref{tab:algebra_packages} shows the different libraries, and what functions they support.
The Chow Patel ILU0 indicates an iterative, highly parallel decomposition\cite{chow_patel_decomp} and application\cite{chow_patel_apply}.

\begin{table}[h]
\centering
\begin{tabular}{c|c|c|c|c|c}
Package & ILU0 & Chow Patel ILU0 & AMG & CPR & bicgstab \\ \hline
cusparse   & \checkmark & - & - & - & \checkmark$^1$ \\
magma      & \checkmark & \checkmark & - & - & \checkmark \\
viennacl   & \checkmark & \checkmark & \checkmark & - & \checkmark \\
ginkgo     & \checkmark & \checkmark & \checkmark & - & \checkmark \\
amgcl      & - & \checkmark & \checkmark & \checkmark & \checkmark \\
rocalution & \checkmark & - & \checkmark & - & \checkmark \\
rocsparse  & \checkmark & - & - & - & \checkmark$^1$ \\
\end{tabular}
\caption{Different sparse linear algebra libraries and their GPU components. $^1$ bicgstab is not readily available but can be constructed using functions from that library.}
\label{tab:algebra_packages}
\end{table}


\section{Implementation}
\label{sec:implementation}

In this chapter, we present the design and implementation of the different preconditioners and solvers that we developed and we provide details about how we integrated them and other third-party GPU libraries into OPM Flow. 

\subsection{OPM Flow}
OPM Flow is composed of two main parts, the assembly and the linear solve. The linear solve is performed by the dune-istl\cite{dune-istl} iterative solver template library. 
OPM uses an ILU0 preconditioned BiCGStab solver as the default configuration. 
When the wells are added to the matrix, it performs the standard bicgstab algorithm.
When the wells are separate, the linear operation (spmv) in the bicgstab is replaced by a combination of an spmv and an operation to apply the wells.
This operation is defined in opm-simulators.
Dune-istl is able to handle blocks of all sizes.

During initialization, the bridge verifies that the implementation specified on the command line is available, and creates the chosen backend solver.

Right before dune-istl is called, our bridge tries to solve the linear system.
At this point, the linear system is stored in the dune-istl BCRS matrix and block vector format.
The block vector internally has a contiguous array, storing all the values.
This is easily copied to the GPU.
The matrix has three components: nonzero values, row pointers, and column indices.
The nonzeroes are blocked, Figure \ref{fig:amgcl_layout} shows how they are stored in memory.
Due to the construction of the matrix in opm-simulators, the nonzeroes are stored contiguously.
These are also easily copied to the GPU, but a check must be made to ensure they are indeed contiguous.
The row pointers and column indices are not readily available.
To get them, the matrix is iterated through, and the sparsity pattern is written to two contiguous arrays.

The linear system is solved via a preconditioned bicgstab solver, capable of handling 3x3 blocks.
If the backend solver is unsuccessful, the call to dune-istl is still made, see subsection \ref{sec:fallback}.

\subsection{Bicgstab solver}
The bicgstab solver needs some basic functions like spmv, axpy, norm, and dot.
The spmv implementation is described in \ref{lab:spmv}.

For the norm and dot, the GPU kernel only returns partial sums, which are added on the CPU.
Each work item calculates the result of one row, and the workgroup reduces this to one value in local memory.
The function $work\_group\_reduce\_add()$ from Opencl 2.0 was not available, since Opencl 1.2 needs to be supported for NVIDIA GPUs.

The default stopping condition in Dune is a relative reduction in error of $0.01$, with a maximum number of linear iterations of $200$.
Our implementation allows us to pass the same stopping criteria, with the same default values.

\subsection{Sparse Matrix-Vector multiplication}
\label{lab:spmv}
A sparse matrix-vector multiplication (spmv) multiplies a sparse matrix A with a dense vector x, resulting in a dense vector y.
Each row i of A can be seen as a sparse vector, and then the inner product between row i and vector x can be calculated to form entry $y_i$.
Every element of y can be calculated in parallel since they do not have dependencies.
For OPM, the elements of A are actually small, dense blocks of size $N\ times N$, and elements of $x$ and $y$ are small, dense vectors of size $N$.
The sparse inner product becomes more complex: the two scalar elements $a_{ij}$ and $x_j$ are now a dense matrix and a dense vector.
To multiply, a standard dense matrix-vector multiplication is performed.

To implement this on the GPU, Algorithm 1 from \cite{blocked_spmv} is used:
Each warp or workgroup is assigned to one or more blockrows.
The warp then iterates through them until all are processed.
For a particular blockrow, the warp covers 32/bs$^2$ blocks at a time, where bs is the block size.
For a block size of 3, the warp covers 3 blocks with 27 threads or workitems, and 5 are left idle.
The thread's position inside a block does not change, it only moves to the same position in another block.
Each thread has its own running sum, and row (r) and column (c) inside the block.
It multiplies its element from the block with the corresponding element of $x$, adding it to the running sum.
It then moves to the next block, which is actually $32/bs^2$ blocks over.
Once the warp has iterated through all blocks in the row, threads that have the same r in the block reduce their running sums into one combined sum, which is written to the output vector.
The paper assumes the values inside the block are stored column-major, but the blocks in OPM are generated by Dune\cite{dune}, which are row-major. This does not change the way the algorithm works, during reading the element of A, the row and column of a thread are simply swapped.

\subsection{ILU0}
ILU0 has two phases: decomposition/creation and application.
During the decomposition, Incomplete LU factorization without fill-in is performed.
To apply the preconditioner, two triangular solves have to be performed with the resulting L and U factors.
Since the sparsity pattern does not change throughout the simulation, the information derived from it can be reused.
The sparsity pattern dictates the amount of parallelism that can be extracted from it.
There are two implemented options to extract parallelism: level scheduling (LS) and graph coloring (GS).
For debugging purposes, performing the ILU sequentially is also an option.
Level scheduling respects indirect dependencies, whereas graph coloring is more aggressive and does not.
If row C depends on row B, and row B depends on row A, then row C indirectly depends on row A.
With LS, row C will be processed after row A is done.
With GC, these two rows will be assigned the same color, and processed simultaneously.
All rows that are processed in parallel are in the same level or color.

After extracting the parallelism, a reordered copy of the matrix is made.
In here, the rows of the matrix are reordered, such that all rows in the same color are in contiguous memory.
The column indices are reordered accordingly.
The $b$ is also reordered.
To perform the GPU kernels, the technique in \cite{blocked_spmv} was used.
This paper describes an spmv operation, but this can also be used for the ILU0 decomposition and application.
During execution of the decomposition and application, each color is processed by a new kernel launch.

\subsection{Block-Jacobi-ILU0}
\label{lab:blockjacobi}
ILU0 is a powerful preconditioner and an important part of the linear solver.
A drawback is that it is fairly slow, a large amount of time is spent in the decomposition and application phases, especially on GPU.
Other parts of the linear solver can be parallelized relatively easily, but ILU0 is an inherently sequential algorithm because the rows depend on each other.
This can be solved by using LS or GC, but the ILU part of the solver can still be the bottleneck.

To further increase parallelism, dependencies can be removed from the decomposition matrix.
The original matrix should still be used for the rest of the solver.
Removing blocks from the matrix for the preconditioner should be done in a smart way, to reduce the quality of the preconditioner as little as possible, while still improving parallelism.
When using N MPI processes to run Flow, the grid can be partitioned into N partitions.
Each partition will be assembled and solved in parallel, with some extra computation and synchronization for the boundary between the partitions.
The partitions are created according to the transmissibilities between cells, these indicate how tightly the cells are coupled.
More information can be found in \cite{Thune_2021}.

The GPU implementation of the Block-Jacobi-ILU0 preconditioner uses the same partitioning algorithm, but the number of partitions is set by the user.
Since every row in the matrix corresponds to an active cell in the grid, it is easy to remove blocks that represent a connection between two cells belonging to different partitions.
These functions are inspired by the work in \cite{andreas}.
This new matrix is then used for the ILU preconditioner.
The sparsity pattern is analyzed, and since it contains fewer blocks, it can extract more parallelism.
The convergence of the ILU preconditioner is not affected too much, if the number of partitions is low enough, and if the partitioning is done in a smart way (using the transmissibilities).

The jacobi matrix has the same number of rows and columns, but less nonzero blocks. To allow the matrix nonzeroes to be copied by a single GPU copy, they must be stored in contiguous memory. The Dune::BCRSMatrix allocates contiguous memory when the maximum number of nonzeroes is passed to the constructor. The number of nonzeroes in the full matrix serves as this upper bound.
For every linear solve, blocks must be copied from the full matrix to the Jacobi matrix.
This can be done by iterating the two matrices simultaneously, and copying a single block when a match is found in the sparsity pattern.
Alternatively, since the nonzeroes of both matrices are stored in contiguous memory, the indices of the blocks that need to be copied can be stored in a big vector $indices$. Each nonzero block gets its own entry in that vector. Copying then occurs in a single loop:
\begin{algorithm}
\caption{Copy indexes}
for i in indices: \\
    jacobi\_matrix\_nnzs[i] = full\_matrix\_nnzs[indices[i]]
\end{algorithm}

\subsection{Well Contributions}
The default CPU implementation does not actually perform a standard bicgstab solver.
After the spmv operation, a separate function is performed to incorporate the contributions of all active wells.
A runtime parameter exists to put these contributions inside matrix A itself, allowing a standard bicgstab solver to be used.
This will increase the complexity of the matrix and might deteriorate converge performance.
This function is described in Algorithm \ref{alg:wells_coupled}.
When the wellcontributions are kept separate, they are added like in Algorithm \ref{alg:wells_separate}.
The spmv from the bicgstab solver is put in a comment.

\begin{algorithm}
\caption{Adding the wellcontributions to matrix A}
for well in wells: \\
\Indp A = A - well.C$^T$ * well.D$^{-1}$ * well.B \\
\label{alg:wells_coupled}
\end{algorithm}

\begin{algorithm}
\caption{Adding the wellcontributions separately}
// y = A * x \\
for well in wells: \\
\Indp 
t1 = well.B * x \\
t2 = well.D$^{-1}$ * t1 \\
y = y - well.C$^{T}$ * t2 \\
\label{alg:wells_separate}
\end{algorithm}

For standardwells:
\begin{itemize}
\item D is a single block
\item B and C have 1 row, with a nonzero block at (0,j) if this well has a perforation at cell j
\end{itemize}

For multisegmentwells:
\begin{itemize}
\item nseg is the number of segments of that well
\item D is a (nseg x nseg) dense block matrix
\item B and C are (nseg x ncells) sparse block matrices, they have a block at (i, j) if this well has a perforation at cell j connected to segment i. The columns of B/C have no more than one nonzero block.
\end{itemize}

To allow separate wellcontributions for standardwells on the GPU, manual kernels have been written to apply them.
The variables B, C, and D can be interpreted as sparse vectors since they have only one row.
There are three contiguous arrays, which store the B, C, and D of all the wells.
For D, the inverse D$^{-1}$ is stored instead, since multiplying with the inverse is easier.
These three arrays are copied to the GPU.
The CUDA and OpenCL kernels for standardwells apply to all wells in parallel.
For every well, one workgroup is launched.

The multisegmentwells are more complex, they have more, irregular-sized data, and are applied using UMFPack\cite{umfpack} on the CPU.
Their application is not implemented on the GPU, instead, the x and y vectors are copied to the CPU, where they are applied with UMFPack.
The resulting y vector is then copied back to the GPU.

\subsection{Amgcl}
Amgcl\cite{amgcl1}\cite{amgcl2} is a header-only C++ library for solving (blocked) sparse linear systems.
It features different preconditioners and iterative solvers.
Different backends like OpenMP, CUDA, or OpenCL can be used.
All these different configurations can be chosen at runtime.
It can also connect to VexCL\cite{vexcl}, ViennaCL, Eigen\cite{eigen} and Blaze\cite{blaze}.

The memory layout for blocked matrices in amgcl is different from OPM, as described in Figure \ref{fig:amgcl_layout}.
This means for every linear solve, the matrix must be transformed to be used by amgcl.

\begin{figure}
\centering
\begin{tabular}{cccccc|cccccc}
0 & 1 & 2 & 9 & 10 & 11     &   0 & 1 & 2 & 3 & 4 & 5 \\
3 & 4 & 5 & 12 & 13 & 14     &  6 & 7 & 8 & 9 & 10 & 11 \\
6 & 7 & 8 & 15 & 16 & 17     &  12 & 13 & 14 & 15 & 16 & 17 \\
\end{tabular}
\caption{Memory layout of blocks in OPM (left) and amgcl (right).}
\label{fig:amgcl_layout}
\end{figure}

To use amgcl, the wellcontributions must be included in the matrix.

\subsection{Rocalution}
Rocalution\cite{rocalution} is a sparse linear algebra library.
It focuses on utilizing fine-grained parallelism and is built on top of AMD's ROCm\cite{rocm} ecosystem.
Numerous iterative solvers, preconditioners, and sparse matrix formats are supported.
It is designed to be able to run on both NVIDIA and AMD GPUs.

The memory format for blocked matrices is row-major, like Dune's BCRSMatrix.
However, the nonzero values inside the blocks must be column-major.
This means that for every linear solve, every matrix block must be transposed before sending it to rocalution.

It does have a BlockJacobi preconditioner, similar to the one described in Subsection \ref{lab:blockjacobi}.
However, it requires the use of GlobalMatrix, a distributed matrix that does not have an easy way to handle blocked matrices.
It also doesn't have a way to determine the blocks manually.
Using the transmissibilities to partition helps to reduce the loss in convergence power.

To use rocalution, the wellcontributions must be included in the matrix.
It might be possible to extend the Operator class, and both apply spmv and add the wellcontributions in its apply() function, but this might break the bicgstab solver and is not tried yet.

\subsection{cuSPARSE}
cuSPARSE\cite{cusparse} is a library with basic linear algebra functions, used for handling sparse matrices.
It is created by NVIDIA and is implemented on top of the CUDA runtime.
Since the bicgstab solver also uses some dense vector functions, cuBLAS\cite{cublas} is used in combination with cuSPARSE to create cusparseSolver.

A CUDA kernel is written to apply the standardwell contributions for cusparseSolver, so that including them in the matrix is not required.
Multisegment wells are still applied on CPU.

\subsection{Rocsparse}
Rocsparse\cite{rocsparse} is the sparse linear algebra implementation of the ROCm framework.
It has the same interface as cusparse.
Using the building blocks provided in the library, and the dense vector functions from rocblas, a simple ILU0-BiCGStab solver can be constructed.
The advantage of using rocsparse over rocalution is that the Block-Jacobi-ILU0 preconditioner can be used.
It should also be easier to allow for separate wellcontributions applications.

\subsection{Fallback}
\label{sec:fallback}
If the chosen BdaSolver fails to converge within the specified limits (default 200 linear iterations), the simulation will fall back to the CPU Dune implementation.
The ILU0 decomposition is performed, and the solver is started.

The ILU0 decomposition from the chosen BdaSolver might be able to be reused instead since it is already performed.
Only the copying added time and could be faster than the CPU ILU0 decomposition.
Since fallbacks are relatively rare, this is not implemented.

\begin{figure}[h!]
\centering
\includegraphics[width=\textwidth]{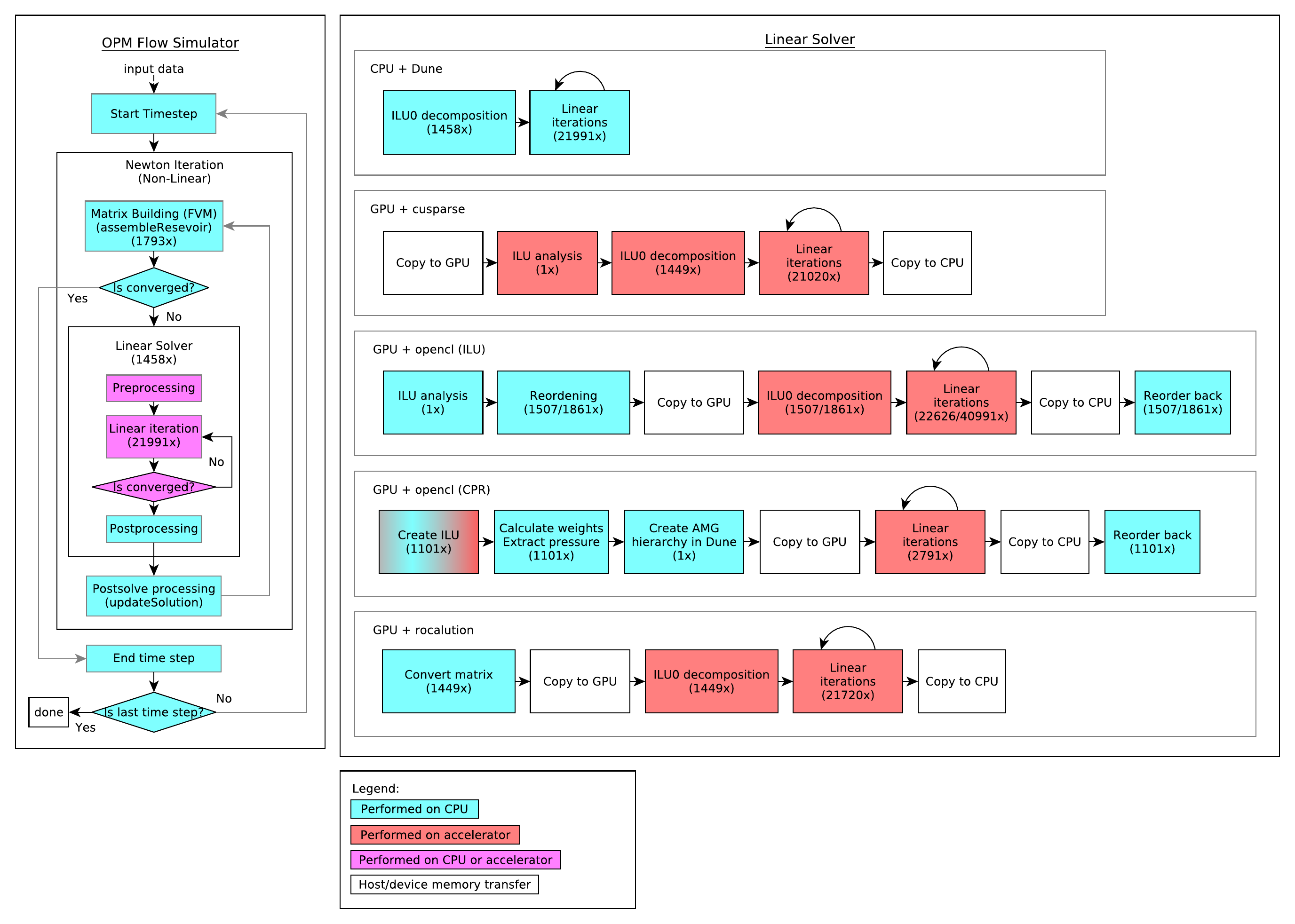}
\label{fig:flow_chart2}
\caption{A flowchart depicting different implementations.}
\end{figure}

\section{Experimental Results}
\label{sec:experimentalresults}

In this section, we present the experimental results. We highlight first the benchmarks used describing their characteristics and introduce the chosen metric format that we will use to report the performance results. Second, we provide details of the different experimental setups that we used in this work including the used CPU and GPU hardware. Finally, after the complete set of performance numbers is presented, we end the section with a discussion about the bottlenecks we encountered explaining potential solutions or future steps to remove those bottlenecks.  

\subsection{Benchmarks}

We use three different benchmarks to perform the evaluations of the different libraries implemented and integrated into OPM. The first use case is NORNE\cite{norne} case, a real-world reservoir off the coast of Norway. NORNE contains 44431 active cells, and the resulting matrix has 320419 blocks when the well contributions are included, or 313133 when they are kept separate. The dataset is open-source and available in the opm-data or opm-tests repository. The second use case is based on a refined version of NORNE, in which the grid size is increased so that it contains 355407 active cells. Finally, the last benchmark used is a proprietary closed-source model containing more than 1 million active cells. Table \ref{table:benchmarks} summarizes the benchmarks providing more details about their characteristics.

\begin{table}[h!]
\centering
\begin{tabular}{c|c|c|c|c|c|c}
Use Case & Active cells & Block Rows (N) & \#NNZs & \#NNZs/row & \#Wells & Type Wells\\ \hline
NORNE  & 44431 &  44431 & 320419 & 7.21 & 3 & STD \\
NORNE refined  & 355407 &  355407 & 2530093 & 7.11 & 3 & STD \\
\textit{BigModel} & 1092723 &  1092723 & 6840701 & 6.26 & 5 & MSW \\
\end{tabular}
\caption{Selected Benchmarks and their Characteristics.}
\label{table:benchmarks}
\end{table}

OPM Flow reports results at the end of each simulation by printing them in the console. The multi-line verbose run-times output of flow is shortened for space reasons in this paper and is summarized below such that one run only takes one line in a performance result table. The defined format is shown in Figure \ref{fig:notation}. Furthermore, when a certain date is mentioned, the latest merge commit before that date/time is used to perform the run.

\vspace{10pt}
\begin{adjustbox}{minipage=\textwidth-2\fboxsep-2\fboxrule, fbox, captionbelow={Notation of results}, label={fig:notation}, nofloat}
Notation: \\
A (B+C+D+E), F, G, H $|$ I, J \\
A: Total time (seconds) \\
B: Assembly time (seconds) \\
C: Linear solve time (seconds) \\
D: Update time (seconds) \\
E: Pre/post step (seconds) \\
F: Overall Linearizations \\
G: Overall Newton Iterations \\
H: Overall Linear Iterations \\
I: Number of colors \\
J: Number of times BdaSolver failed and used OPM solver instead \\
If only 1 number after the '$|$' is reported, it is the number of failed solves.
\end{adjustbox}
\vspace{10pt}

\subsection{Experimental Setup}

Simula\cite{simula1} is a Norse research institute. Its main activities are research, innovation, and education. The research is conducted in five areas: communication systems, cryptography, machine learning, scientific computing, and software engineering. Simula hosts the Experimental Infrastructure for Exploration of Exascale Computing\cite{simula2}\cite{simula3}\cite{simula4} (eX$^3$), a high-performance computing cluster. Some of its nodes are listed in Table \ref{tab:simula_nodes}, while Table \ref{tab:gpus} highlights the specifications of the different hardware available in terms of the number of cores, maximum tera floating point operations per second (TFLOPS), maximum available memory on board, and the maximum bandwidth on the accelerator card. 

\begin{table}[h!]
\centering
\begin{tabular}{l|c|c}
Name      & CPU & GPU \\ \hline
g001      & Intel Xeon Platinum 8168 @ 2.7 GHz & NVIDIA V100-SXM3-32GB \\
g002      & AMD EPYC 7763 & NVIDIA A100-SXM-80GB \\
n013/n014 & AMD EPYC 7763 & NVIDIA A100-SXM-40GB \\
n004      & AMD EPYC 7601 & AMD Instinct MI100 \\
n015/n016 & AMD EPYC 7763 & 2x AMD Instinct MI210 \\
\end{tabular}
\caption{The Different Simula Nodes and their Hardware Configuration.}
\label{tab:simula_nodes}
\end{table}

\begin{table}[h!]
\centering
\begin{tabular}{c|c|c|c|c}
\multirow{2}{*}{Name}      & \multirow{2}{*}{Num. cores$^1$} & \multirow{2}{*}{Max. FP64 TFLOPS} & Memory & Memory \\
&&& size (GB) & bandwidth (GB/s) \\ \hline
NVIDIA Tesla V100  & 5120 &  7.8 & 32 & 900 \\
NVIDIA Tesla A100  & 6912 &  9.7 & 40/80 & 2000 \\
AMD Instinct MI100 & 7680 & 11.5 & 32 & 1229 \\
AMD Instinct MI210 & 6656 & 22.6 & 64 & 1638 \\
AMD Instinct MI250 & 13312 & 45.3 & 128 & 3277 \\
\end{tabular}
\caption{Different GPUs and their Specifications. $^1$ NVIDIA CUDA Cores are not the same as AMD Stream Processors.}
\label{tab:gpus}
\end{table}

\subsection{Performance Results}

First of all, to compare how the new hardware from Simula performs against our previous work, to compare the hardware from \cite{own_paper} and the nodes from Simula, the release/2020.10-rc4 is benchmarked again in Table \ref{tab:simula_2020}. All four configurations are slower on the g001 node. This is surprising since the g001 node has a CPU with a faster base clock and a faster boost clock. Their GPUs have a different memory size, but the NORNE test case does not use enough memory to fill either. One possible explanation is that the Xeon E5-2698v4 has a cache of 50 MB, whereas the Platinum 8168 has 33 MB.

\begin{table}[h!]
\centering
\begin{tabular}{c|c|l}
\multirow{4}*{coupled}
& none      & 571 (152+319+63), 1793, 1458, 21196 \\
& cusparse  & 371 (151+123+62), 1783, 1449, 21010 $|$ 0 \\
& opencl LS & 559 (156+303+64), 1844, 1507, 22626 $|$ 0 \\
& opencl GC & 510 (180+218+76), 2109, 1774, 40863 $|$ 0 \\
\end{tabular}
\caption{g001, Tesla V100, release/2020.10-rc4}
\label{tab:simula_2020}
\end{table}

\begin{table}[h!]
\centering
\begin{tabular}{c|c|c}
Type Well / Library & STDWELL & MSWELL \\\hline
Dune & yes & yes \\ 
cusparse & yes (GPU) & yes (CPU) \\ 
opencl & yes (GPU) & yes (CPU) \\ 
rocsparse & yes (GPU) & yes (CPU) \\ 
rocalution & no & no \\ 
amgcl & no & no \\ 
\end{tabular}
\caption{Summary of How the Types of Wells and their Contributions to the System Matrix are Supported by the OPM Flow (GPU) Backends.}
\label{tab:wells}
\end{table}

Please note that previously, we experimented only with a cusparse and OpenCL solvers using solely an NVIDIA GPU and using a linear solver that allowed only for coupled wellcontributions. In the following, we test multiple solvers using other libraries (i.e., rocalution, rocsparse, and amgcl), we add an optimization technique for the ILU0 preconditioner, we extend the experimental results to include also AMD GPU hardware, and finally, we develop support for both coupled and separate for the well contributions as well as different types of wells in the GPU linear solvers. Nevertheless, because of the limited flexibility to modify a library linear solver, the separate well contributions are not fully supported for all the libraries, as explained in Section \ref{sec:implementation} Implementation. Table \ref{tab:wells} summarizes the type of well (standard (STD) or multisegment (MS)) and its support in the different GPU libraries. 

Table \ref{tab:lin_solver_times} highlights the different solver libraries we evaluated and that we run on different CPU and GPU hardware. For space reasons, we highlight only the linear solver time and do not report the total time it took to run a complete reservoir simulation with OPM flow. However, the complete results of the runs are shown in Tables \ref{tab:norne_times}, \ref{tab:norne_ref_times}, and \ref{tab:bigmod_times} for the three use cases, respectively. Please note that due to space reasons, we limit the presentation only to the fastest results that were achieved, which for the GPU was the ROCm-based implementation, and we compare it against the multiple-rank scaled results using the DUNE library on the CPU.

\begin{table}[h!]
\centering
\begin{tabular}{c|c|c|c|c|c|c|c}
& Benchmark: & \multicolumn{2}{|c|}{NORNE} &\multicolumn{2}{|c|}{NORNE Refined} & \multicolumn{2}{|c}{Big Model} \\\hline
Library & Hardware & \multirow{2}{*}{Coupled} & \multirow{2}{*}{Separate} & \multirow{2}{*}{Coupled} & \multirow{2}{*}{Separate} & \multirow{2}{*}{Coupled} & \multirow{2}{*}{Separate} \\ 
Configuration & Device(s) & & & & & & \\\hline\hline
DUNE,   1 MPI & EPYC 7763 & 174 & 178 & 3832 & 3847 & \textit{Err2} & 3758\\
DUNE,   2 MPI & EPYC 7763 &  86 &  82 & 2161 & 2142 & \textit{Err2} & 1853\\
DUNE,   4 MPI & EPYC 7763 &  45 &  39 & 1283 & 1162 & \textit{Err2} & 1097\\
DUNE,   8 MPI & EPYC 7763 &  31 &  26 &  696 &  618 & \textit{Err2} & 593\\
DUNE,  16 MPI & EPYC 7763 &  19 &  17 &  424 &  329 & \textit{Err2} & 359\\
DUNE,  32 MPI & EPYC 7763 &  19 &  17 &  291 &  255 & \textit{Err2} & 251\\
DUNE,  64 MPI & EPYC 7763 & \textbf{16} & \textbf{14} & \textbf{267} & \textbf{212} & \textit{Err2} & \textbf{215}\\
DUNE, 128 MPI & EPYC 7763 & \textit{Err1} & \textit{Err1} & 341 & 277 & \textit{Err2} & 281\\
rocsparse-0   & MI210 &  95 & 91 & 779 & 873 & \textbf{908} & \textbf{804}\\
rocsparse-150 & MI210 &  \textbf{53} & 58 & \textbf{685} & 867 & 1030 & 995 \\
rocalution    & MI210 &  94 & N/A & 1014 & N/A & \textit{same rocs} & N/A\\
opencl-0 LS   & MI210 & 384 & 506 & 2096 & 2685 & \textit{slow}& \textit{slow}\\
opencl-150 LS & MI210 & 101 & 147 & 1440 & 1289 & \textit{slow}& \textit{slow}\\
amgcl, vexcl & MI210 & 235 & N/A & 4729 & N/A & \textit{slow} & N/A \\
amgcl, vexcl & V100 & 276 & N/A & 7645 & N/A & \textit{slow} & N/A \\
amgcl, vexcl & A100 & 195 & N/A & 4419 & N/A & \textit{slow} & N/A \\
amgcl, cuda  & V100 & 685 & N/A & Err & N/A & \textit{slow} & N/A \\
amgcl, cuda  & A100 & 772 & N/A & Err & N/A & \textit{slow} & N/A \\
cusparse-0    & V100 & 95 & 86 & 1205 & 913 & \textit{same rocs} & \textit{same rocs} \\
cusparse-0    & A100 & 118 & 101 & 1025 & 976 & \textit{same rocs} & \textit{same rocs} \\
cusparse-150  & V100 &  60 & 59 & 1103 & 1238 & \textit{same rocs} & \textit{same rocs} \\
cusparse-150  & A100 &  57 & \textbf{51} & 807 & \textbf{845} & \textit{same rocs} & \textit{same rocs} \\
rocsparse-0   & MI100& 101 & 108 & 950 & 1094 & \textit{same rocs} & \textit{same rocs} \\
rocsparse-150 & MI100&  55 &  71 & 964 & 1170 & \textit{same rocs} & \textit{same rocs} \\
rocalution    & MI100& 110 & N/A & 1211 & N/A & \textit{slow}& N/A\\
opencl-0 LS   & MI100& 429 & 254 & 3850 & 3465 & \textit{slow} & \textit{slow} \\
opencl-150 LS & MI100& 143 & 135 & 1753 & 1867 & \textit{slow} & \textit{slow} \\
amgcl, vexcl 7& MI100& 241 & N/A & 6006 & N/A & \textit{slow} & N/A \\
\end{tabular}
\caption{ OPM Flow Linear Solver Time in Seconds Running on Different Hardware Devices Using the Masters from 2023-4-13 Configured with the Default ILU0 Preconditioned BiCGStab Solver.}
\label{tab:lin_solver_times}
\end{table}

where, 

\begin{itemize}
    \item \textit{N/A} means the option is not available due to separate wells not being supported for that library,
    \item \textit{Err1} is an OPM flow error related to the study case being too small to be distributed on 128 MPI processes,
    \item \textit{Err2} OPM flow did not converge when using the default setting of chopping the time step ten times.
    \item \textit{Slow} means the run was stopped before finishing due to taking a longer time compared to the software 1 MPI process, and 
    \item \textit{Same} means the run matches the mentioned library.
\end{itemize}

Analyzing the results, we find that a GPU-based ILU0-preconditioned BiCGStab linear solver can accelerate a single dual-threaded MPI process up to 5.6x times (3832s vs 685s) for a medium-size reservoir model and up to 3.3x (174s vs. 53s) for a small size model. For the medium-size model, i.e., NORNE refined, the performance is equivalent to approximately 8 MPI dual-threaded processes when using the same type of preconditioned solver. Therefore, we believe that when benchmarking a small-size model such as NORNE, which allows the model data to fit (almost) entirely into the CPU L1 caches, the benefits of using a GPU accelerator are minimal as highlighted by the experimental results. 

Comparing the AMD ROCm rocsparse library solver with the NVIDIA CUDA corresponding implementation, we find that both perform equally well. Similarly, when we compare the older with the newer generation hardware devices, we notice an almost equal improvement in performance out-of-the-box, namely 22\% for AMD in favor of the MI210 vs. MI100 and 15\% for NVIDIA A100 vs. V100. When we analyze our fully manual implementation that uses manually written OpenCL kernel functions, which do not benefit from the highly optimized assembly-like kernel primitives, we notice a substantial slowdown of 4x (384s vs. 95s) against both the AMD ROCm rocsparse library without Jacobi and the rocalution libraries. However, the benefit of relaxing the ILU0 preconditioner by enhancing it with the Jacobi-based preprocessing and, hence, increasing the parallelism, is the highest in the case of the manual OpenCL implementation with a speedup of 3.8x (384s vs. 101s) when running on the more recent AMD hardware. Finally, we also evaluated the third-party open-source library amgcl and the experiments with an out-of-the-box configuration revealed a surprising 1.35x slowdown (235s vs 174s) when compared to a single dual-threaded CPU run of flow, or in other words, a speedup of 2.5x (235s vs. 94s) in favor of the AMD ROCm rocalution based solver. Please note that the aforementioned evaluation is considering only the NORNE benchmark. However, an equivalent analysis can be performed for the other two use cases with approximately similar conclusions. This is left as an exercise for the reader because this can be easily inferred from Table \ref{tab:lin_solver_times}. 

Tables \ref{tab:norne_times} to \ref{tab:bigmod_times} show the complete performance results of a complete OPM flow reservoir simulation, including the times for the assembly, system update, and the pre- and post-processing time, as well as the number of iterations performed in the linear solver following the format described in Figure \ref{fig:notation}. Due to space reasons, we highlight in these tables only the runs with different MPI processes and the AMD ROCm rocsparse solvers that performed the best among the GPU solutions tested. Please note that the GPU implementations do not currently support multiple ranks and, as such, cannot benefit from parallel multi-rank assembly, update, and post-processing that further improves the performance of multiple CPU MPI ranks experiments vs the GPU ones because the assembly is performed on a single CPU when a GPU solver is chosen. 

\begin{table}[h!]
\centering
\begin{tabular}{c|c|r}
\multirow{7}*{coupled}
& DUNE, 1 MPI & 388 (65+174+76+69), 2041, 1660, 22346 \\
& DUNE, 4 MPI & 139 (25+ 45+24+27), 1989, 1610, 22305 \\
& DUNE, 8 MPI & 113 (17+ 31+22+21), 1936, 1596, 21900 \\
& DUNE, 16 MPI & 84 (13+ 19+ 6+16), 1909, 1539, 22056 \\
& DUNE, 64 MPI &  78 ( 8+ 16+ 3+14), 1916, 1548, 23318 \\
& rocsparse-0  & 362 (85+95+98+80), 1946, 1578, 21836 \\
& rocsparse-150  & 330 (93+47+96+89), 1894, 1528, 25575  \\
\hline
\multirow{7}*{separate}
& DUNE, 1 MPI & 438 (77+178+102+77), 1894, 1526, 21920 \\
& DUNE, 4 MPI & 125 (21+ 39 + 21+25), 1883, 1519, 22192 \\
& DUNE, 8 MPI & 129 (26+ 26 + 29+23), 1894, 1530, 22653 \\
& DUNE, 16 MPI & 85 ( 8 + 17  +  8 +16), 1882, 1519, 23026 \\
& DUNE, 64 MPI & 81 (13 + 14+ 3 +15), 1913, 1544, 23815 \\
& rocsparse-0  & 375 (95 +91+100+83), 1951, 1580, 23049 \\
& rocsparse-150  & 359 (101+58+103+90), 1953, 1582, 27014 \\
\end{tabular}
\caption{\textit{NORNE} benchmark, EPYC 7763 CPU vs. AMD Instinct MI210, masters of 2023-4-12 10:00}
\label{tab:norne_times}
\end{table}

\begin{table}[h!]
\centering
\begin{tabular}{c|c|r}
\multirow{7}*{coupled}
& DUNE, 1 MPI & 7015 (992+3832+1344+826), 3933, 3422, 65295 \\
& DUNE, 4 MPI & 2272 (330+1283+ 365+244), 3861, 3345, 64765 \\
& DUNE, 8 MPI & 1286 (215+ 696+ 156+138), 3915, 3392, 64997 \\
& DUNE, 16 MPI &   885 ( 124+ 424+ 108+ 90), 4058, 3536, 67293 \\
& DUNE, 64 MPI & 610 ( 77 + 267 + 36 + 50), 3799, 3289, 66289 \\
& rocsparse-0  & 4700 (1238+779+1716+944), 3905, 3387, 65494 \\
& rocsparse-150  & 5077 (1315+685+2070+982), 4011, 3493, 69223 \\
\hline
\multirow{7}*{separate}
& DUNE, 1 MPI & 7019 (981+3847+1334+838), 3936, 3419, 66986 \\
& DUNE, 4 MPI & 2153 (331+1162+ 366+240), 3823, 3316, 67790 \\
& DUNE, 8 MPI & 1198 (196+ 618+ 160+132), 3865, 3344, 66264 \\
& DUNE, 16 MPI &   774 ( 109+ 392+ 117+ 85), 3678, 3176, 64504 \\
& DUNE, 64 MPI & 562 ( 75 + 212 + 33 + 49), 3799, 3292, 68168 \\
& rocsparse-0  & 5176 (1470+873+1805+1000), 4532, 3963, 82683 \\
& rocsparse-150  & 4943 (1359+867+1755+936), 4495, 3940, 85263 \\
\end{tabular}
\caption{\textit{NORNE refined} benchmark, EPYC 7763 CPU vs. AMD Instinct MI210, masters of 2023-4-12 10:00}
\label{tab:norne_ref_times}
\end{table}

\begin{table}[h!]
\centering
\begin{tabular}{c|c|r}
\multirow{7}*{separate}
& DUNE, 1 MPI & 9150 (2209+3755+2859+294), 2763, 2465, 17419 \\
& DUNE, 4 MPI & 2513 ( 623+1097+ 601 +132), 2802, 2514, 17648 \\
& DUNE, 8 MPI & 1390 ( 333+ 593 + 308 +103), 2834, 2547, 18144 \\
& DUNE, 16 MPI & 888 ( 219+ 359 + 161 + 98), 2802, 2510, 17942 \\
& DUNE, 64 MPI & 528 ( 125 + 215 + 50 + 78), 2675, 2384, 18104 \\
& rocsparse-0  & 6696 (2738+ 804+2816+298), 2666, 2379, 17621 \\
& rocsparse-150  & 6156 (2462+ 955+2380+282), 2782, 2488, 17244 \\
\end{tabular}
\caption{\textit{Big model} benchmark, EPYC 7763 CPU vs. AMD Instinct MI210, masters of 2023-4-12 10:00}
\label{tab:bigmod_times}
\end{table}

\subsection{Profile numbers}

To understand the bottlenecks of the ILU0-preconditioned linear solver on the GPU, we perform several profiles using the profilers available in the different GPU vendors' toolboxes. That is, we use the NVIDIA Nsight Compute (ncu) profiler to gain insight into the OPM flow simulator for the fastest CUDA implementation, namely the cusparse ILU0 solver with block Jacobi, and run it with settings for 0 and 150 Jacobi blocks. We analyze the profile numbers for all the benchmarks for both cases when the well contributions are included in the matrix or treated separately. For the AMD GPU implementation, we select the rocsparse implementation with block Jacobi running with the same configurations for three use cases using the rocprof profilers. Additionally, we profile the big model benchmark using the more high-level omniperf tool from the ROCM framework. The results of these profiles are shown in Tables \ref{tab:ncu_profiler}, \ref{tab:rocprofiler}, and \ref{tab:omniperf} for the three benchmarks. Please note that we do not report numbers for the complete preconditioned solver, but we select only the most important kernels in the preconditioner (i.e., the lower and upper triangular solves) and blocked scale (i.e., bsrsv2), a fully parallel operation, and blocked sparse matrix-vector multiplication (i.e., bsrmv) in the solver part to show the (in)efficiencies of the GPU implementation when scaling the model size.

\begin{table}[h!]
\centering
\begin{tabular}{c|c|c|c|c|c|c|c}
& Benchmark: & \multicolumn{2}{|c|}{NORNE} &\multicolumn{2}{|c|}{NORNE Refined} & \multicolumn{2}{|c}{Big Model} \\\hline
Profiled & Profiled & \multirow{2}{*}{0 blocks} & \multirow{2}{*}{150 blocks} & \multirow{2}{*}{0 blocks} & \multirow{2}{*}{150 blocks} & \multirow{2}{*}{0 blocks} & \multirow{2}{*}{150 blocks} \\ 
Kernel & Metric & & & & & & \\\hline\hline
\multirow{7}{*}{scale\_bsrsv2} & Compute(SM)(\%) & 7 & 7 & 18 & 18 & 22 & 22\\
 & SM Busy (\%) & 15 & 15 & 24 & 24 & 24 & 24 \\
 & Mem BW (GB/s) & 156 & 159 & 550 & 540 & 960 & 964\\
 & Mem Busy(\%) & 15 & 15 & 47 & 46 & 67 & 67\\
 & L1 Hit Rate(\%) & 0 & 0 & 0 & 0 & 0 & 0\\
 & L2 Hit Rate(\%) & 56 & 55 & 52 & 52 & 59 & 59\\
 & Occupancy (\%) & 49 & 49 & 77 & 76 & 81 & 81\\\hline
\multirow{7}{*}{solve\_lower} & Compute(SM)(\%) & 23 & 23 & 27 & 34 & 33 & 41 \\
 & SM Busy (\%) & 19 & 30 & 28 & 40 & 34 & 43\\
 & Mem BW (GB/s) & 11 & 40 & 40 & 76 & 65 & 104\\
 & Mem Busy(\%) & 23 & 18 & 29 & 27 & 30 & 32\\
 & L1 Hit Rate(\%) & 1 & 4 & 2 & 5 & 3 & 5\\
 & L2 Hit Rate(\%) & 72 & 73 & 75 & 73 & 74 & 72\\
 & Occupancy (\%) & 81 & 64 & 81 & 75 & 83 & 82\\\hline
\multirow{7}{*}{solve\_upper} & Compute(SM)(\%) & 23 & 23 & 27 & 34 & 36 & 42\\
 & SM Busy (\%) & 19 & 31 & 28 & 43 & 37 & 45\\
 & Mem BW (GB/s) & 11 & 42 & 39 & 77 & 72 & 109\\
 & Mem Busy(\%) & 23 & 18 & 30 & 27 & 30 & 32\\
 & L1 Hit Rate(\%) & 1 & 4 & 2 & 5 & 3 & 5\\
 & L2 Hit Rate(\%) & 72 & 74 & 76 & 73 & 75 & 72\\
 & Occupancy (\%) & 81 & 66 & 81 & 78 & 83 & 78\\\hline
\multirow{7}{*}{bsrmv} & Compute(SM)(\%) & 44 & 43 & 51 & 51 & 53 & 53\\
 & SM Busy (\%) & 49 & 49 & 52 & 52 & 53 & 53\\
 & Mem BW (GB/s) & 297 & 295 & 364 & 364 & 343 & 343\\
 & Mem Busy(\%) & 53 & 50 & 59 & 59 & 59 & 59\\
 & L1 Hit Rate(\%) & 22 & 22 & 22 & 22 & 21 & 22 \\
 & L2 Hit Rate(\%) & 29 & 28 & 40 & 40 & 42 & 43\\
 & Occupancy (\%) & 56 & 56 & 59 & 59 & 59 & 59\\\hline
\end{tabular}
\caption{OPM Flow Profiles of the GPU cuSparse Linear Solver using the 0 and 150 Jacobi Blocks Preconditioner Settings on the NVIDIA A100 GPU using the NCU Profiler.}
\label{tab:ncu_profiler}
\end{table}

\begin{table}[h!]
\centering
\begin{tabular}{c|c|c|c|c|c|c|c}
& Benchmark: & \multicolumn{2}{|c|}{NORNE} &\multicolumn{2}{|c|}{NORNE Refined} & \multicolumn{2}{|c}{Big Model} \\\hline
Profiled & Profiled & \multirow{2}{*}{0 blocks} & \multirow{2}{*}{150 blocks} & \multirow{2}{*}{0 blocks} & \multirow{2}{*}{150 blocks} & \multirow{2}{*}{0 blocks} & \multirow{2}{*}{150 blocks} \\ 
Kernel & Metric & & & & & & \\\hline\hline
\multirow{10}{*}{bsrilu0\_2\_8} & VALUInsts & 1688 & 700 & 1352 & 849 & 1276 & 1020\\
& SALUInsts   & 1323 & 456 & 1000 & 554 & 936 & 702\\
& VALUUtilization (\%)  & 84 & 72 & 80 & 73 & 80 & 77\\
& VALUBusy (\%)          & 16 & 22 & 20 & 34 & 20 & 24\\
& SALUBusy (\%)          & 12 & 14 & 14 & 21 & 15 & 16\\
& MemUnitBusy (\%)        & 64 & 46 & 68 & 70 & 70 & 70\\
& MemUnitStalled (\%)      & 1 & 3 & 4 & 9 & 7 & 11\\
& L2CacheHit (\%)  & 99 & 97 & 99 & 94 & 98 & 95\\
& WriteSize (KB)  & 20 & 16 & 195 & 199 & 640 & 629\\
& FetchSize (KB)  &  &  &  &  &  & \\\hline
\multirow{10}{*}{bsrsv\_lower} & VALUInsts & 418 & 249 & 300 & 264 & 274 & 260\\
& SALUInsts   & 404 & 120 & 169 & 122 & 140 & 123\\
& VALUUtilization (\%)  & 65 & 49 & 52 & 49 & 50 & 49\\
& VALUBusy (\%)          & 15 & 33 & 34 & 44 & 38 & 45\\
& SALUBusy (\%)          & 15 & 16 & 18 & 19 & 19 & 21\\
& MemUnitBusy (\%)        & 54 & 55 & 71 & 73 & 82 & 81\\
& MemUnitStalled (\%)      & 1 & 3 & 3 & 6 & 12 & 9\\
& L2CacheHit (\%)  & 98 & 86 & 91 & 75 & 84 & 69\\
& WriteSize (KB)  & 1 & 1 & 12 & 28 & 67 & 81\\
& FetchSize (KB)  & 19 & 16 & 160 & 194 & 559 & 722\\\hline
\multirow{10}{*}{bsrsv\_upper} & VALUInsts & 433 & 264 & 309 & 270 & 277 & 264\\
& SALUInsts   & 430 & 148 & 183 & 131 & 143 & 128\\
& VALUUtilization (\%)  & 66 & 52 & 53 & 50 & 51 & 50\\
& VALUBusy (\%)          & 14 & 23 & 33 & 42 & 39 & 47\\
& SALUBusy (\%)          & 14 & 13 & 19 & 20 & 20 & 22\\
& MemUnitBusy (\%)        & 56 & 56 & 71 & 71 & 86 & 80\\
& MemUnitStalled (\%)      & 1 & 4 & 3 & 6 & 14 & 8\\
& L2CacheHit (\%)  & 98 & 88 & 91 & 74 & 83 & 66\\
& WriteSize (KB)  & 1 & 1 & 12 & 28 & 62 & 86\\
& FetchSize (KB)  & 17 & 14 & 156 & 199 &  540 & 725\\\hline
\multirow{10}{*}{bsrxmvn\_3x3} & VALUInsts & 154 & 154 & 141 & 141 & 136 & 136\\
& SALUInsts   & 31 & 31 & 30 & 30 & 29 & 29\\
& VALUUtilization (\%)  & 66 & 66 & 72 & 72 & 70 & 70\\
& VALUBusy (\%)          & 5 & 5 & 6 & 6 & 7 & 7\\
& SALUBusy (\%)          & 1 & 1 & 1 & 1 & 1 & 1\\
& MemUnitBusy (\%)        & 69 & 69 & 93 & 92 & 98 & 98\\
& MemUnitStalled (\%)      & 5 & 6 & 12 & 12 & 14 & 14\\
& L2CacheHit (\%)  & 62 & 62 & 62 & 63 & 64 & 64\\
& WriteSize (KB)  & 1 & 1 & 8 & 8 & 25 & 26\\
& FetchSize (KB)  & 30 & 30 & 230 & 230 & 361 & 632\\\hline
\end{tabular}
\caption{OPM Flow Profiles of the GPU rocSparse Linear Solver using the 0 and 150 Jacobi Blocks Preconditioner Settings on the AMD MI210 using the rocprofiler.}
\label{tab:rocprofiler}
\end{table}

The first thing we notice is the contrast between the scale\_bsrsv2 operation and how this is able to utilize the resources efficiently when increasing the size of the model due to its inherent parallelism as opposed to the lower and upper solver that is the core of the ILU0 application, which are not able to leverage the massive resources available on the GPU. For example, in terms of memory throughput, only 104 GB/s is achieved which accounts for less than 10\% of the available memory bandwidth being utilized. For the spmv kernel, this reaches 343 GB/s which is more than 3x better, but it is still far from the maximum available. This is caused in part due to the sequential nature of the ILU0 application and the irregular memory accesses in the spmv. Furthermore, we see that when increasing the parallelism in the ILU0 preconditioner using the Jacobi technique, the bandwidth is almost doubled from 118 GB/s to 196 GB/s, as shown in Table \ref{tab:omniperf}. However, this is still just a small percentage of the total maximum available bandwidth. Please note that due to issues with running omniperf for NORNE and NORNE modified, we report omniperf profile numbers only for the bigmod use case. In terms of resources, we notice that we have plenty of resources left on the board that are not utilized (especially for the lower and upper solves), which can be attributed to the fact that we do not have enough data to feed these resources since the sequential accesses and dependencies in the ILU0 nature prevent further parallelism.   

\begin{table}[h!]
\centering
\begin{tabular}{c|c|c|c}
& Benchmark: & \multicolumn{2}{|c}{Big Model} \\\hline
Profiled & Profiled & \multirow{2}{*}{0 blocks} & \multirow{2}{*}{150 blocks} \\ 
Kernel & Metric & & \\\hline\hline
\multirow{8}{*}{ilu\_apply} & Bandwidth(GB/s) & 118 & 196 \\
 & LDS util (\%) & 30 & 37 \\
 & Vector L1 Data Cache Hit(\%) & 72 & 71\\
 & Vector L1 Data Cache BW(\%) & 16 & 16\\
 & Vector L1 Data Cache util(\%) & 99 & 98\\
 & Vector L1 Coalescing(\%) & 56 & 51\\
 & L2 Cache Hit (\%) & 84 & 69\\
 & L2 Cache Util (\%) & 99 & 98 \\\hline
\multirow{1}{*}{ilu\_decomp} & Bandwidth(GB/s) & 30 & 57 \\\hline
\multirow{1}{*}{spmv} & Bandwidth(GB/s) & 806 & 807 \\
\end{tabular}
\caption{OPM Flow Profiles of the GPU rocSparse Linear Solver using the 0 and 150 Jacobi Blocks Preconditioner Settings on the AMD MI210 using the Omniperf Tool.}
\label{tab:omniperf}
\end{table}

\begin{table}[h!]
\centering
\begin{tabular}{c|c|c|c|c|c|c|c|c|c}
 & Benchmark: & \multicolumn{2}{|c|}{NORNE} &\multicolumn{2}{|c|}{NORNE Refined} & \multicolumn{4}{|c}{Big Model} \\\cline{2-10}
Runtime & Config Wells & \multicolumn{2}{|c|}{coupled} & \multicolumn{2}{|c|}{coupled} & \multicolumn{2}{|c|}{coupled} & \multicolumn{2}{|c}{separate}\\ \cline{2-10}
metric & \#Jac Blocks & 0 & 150 & 0 & 150 & 0 & 150 & 0 & 150 \\ \hline\hline
\multicolumn{2}{l|}{copy\_to\_jacobiMatrix} & - & 3.4 & - & 55.3 & - & 126.9 & - & 113.7  \\
\multicolumn{2}{l|}{check\_zeros\_on\_diagonal}  & 1.7 & 2.1 & 15.5 & 30.9 & 36.4 & 72.6 & 43.1 & 60.7 \\
\multicolumn{2}{l|}{copy\_to\_GPU}  & 5.7 & 6.7 & 48.1 & 84.1 & 128.7 & 208.5 & 112.2 & 205.5 \\
\multicolumn{2}{l|}{decomp}  & 4.7 & 1.4 & 49.9 & 19.4 & 90.7 & 68.6 & 36.7 & 26.0 \\
\multicolumn{2}{l|}{total\_solve}  & 77.6 & 32.1 & 603.3 & 450.1 & 299.5 & 245.6 & 555.2 & 576.5 \\
\multicolumn{2}{c|}{prec\_apply}  & 66.9 & 22.3 & 532 & 376 & 261.9 & 209.0 & 298.9 & 232.7 \\
\multicolumn{2}{c|}{spmv}  & 2.4 & 2.7 & 41.3 & 43 & 22.2 & 22.4 & 28.7 & 28.1 \\
\multicolumn{2}{c|}{wells}  & - & - & - & - & - & - & 209.1 & 297.1 \\
\multicolumn{2}{c|}{rest}  & 5.8 & 5 & 20.2 & 21 & 9.3 & 8.3 & 11.4& 11.4 \\\hline
\multicolumn{2}{l|}{TOTAL time timers}  & 89.7 & 45.7 & 716.8 & 639.8 & 555.9 & 722.8 & 740.1 & 975.2 \\
\multicolumn{2}{l|}{Linear solve time (OPM)}  & 91.2 & 47.1  & 780 & 671 & 1055.9 & 1304.5 & 781.8 & 1021.9 \\\hline
\multicolumn{2}{l|}{linearize wells}  & 0.6 & 0.6 & 3.9 & 4.1 & 476.7 & 460.3 & 1.8 & 1.7 \\
\end{tabular}
\caption{ OPM Flow Linear Solver Time in Seconds Running on Different Hardware Devices Using the Masters from 2023-4-13.}
\label{tab:runtime_profiles}
\end{table}

Finally, to clearly match the above fine-metrics profile, we perform also course-grained profiles in the form of a run-time analysis by measuring the total time it takes to run the different parts in the GPU solver backend of the OPM flow simulator. We timed the different parts of the rocsparse GPU solver implementation and we printed the accumulating timers at the end of the simulation. Table \ref{tab:runtime_profiles} highlights the output of this analysis. These numbers are grouped by the computationally different parts of the solver, such as the total time it takes to do the decomposition, how much it takes to perform spmv, as well as how long solving for the well contributions takes. Additionally, the copying and transfer transfers from/to the GPU are included in this analysis. We observe that for the latter, these take a non-negligible amount and that they scale poorly with increasing size, which partly explains the lower relative speed-up in performance when comparing NORNE against NORNE refined, i.e., 89.7s vs 45.7s, which is almost 2x speedup, versus 716.8s vs 639.8s that is only a minor 12\% speed-up. The trend is extended for the bigger model, in which case we even notice a slowdown for the more parallel 150-blocks ILU0 implementation. Therefore, one of the key conclusions to optimize is to reduce the cost for the memory transfers and copy to Jacobi matrix. Furthermore, the different type of wells seems to contribute as well to the relative slow-down and it will require attention.

\newpage
\section{Conclusion and Future Work}
\label{sec:conclusion}

In this paper, we have evaluated the potential of using GPUs in the OPM project, which contains open-source implementations of several reservoir simulation CFD applications. To perform this study, we have developed several manual ILU0 preconditioned BiCGStab iterative solvers using OpenCL and CUDA and experimented with both AMD and NVIDIA GPUs. Furthermore, we have developed a bridge to seamlessly connect the GPU solvers and third-party scientific libraries into OPM. Finally, we provided extensive evaluation results for all different solver configurations used running on different GPU hardware offered by different hardware vendors. To perform our bench-marking, we use small, medium, and large use cases, starting with the public test case NORNE that includes approximately 50k active cells, and ending with a large model that includes approximately 1 million active cells. We find that a GPU can accelerate a single dual-thread MPI process up to 5.6 times and that it can compare with around 8 dual-thread MPI processes. Finally, the source codes for the GPU kernels, solvers, and the integration with OPM \textit{Flow} are made available in the git repositories of the OPM project \cite{opmrepo}. 

In the future, we plan to further analyze the potential slowdowns in the GPU solvers that prevented us from obtaining more impressive speedup numbers. First and foremost, we need to continue with the profiling of the kernels and understand if the matrix data can be better encoded to maximize the cache hit and increase the bandwidth utilized. Furthermore, we should also investigate the impact of adding the well contribution to the matrix or the impact of the memory transfers when the contributions are calculated on the CPU. This could clarify why the speedup for the largest model is smaller than for the small-scale NORNE test case. Finally, several optimizations could be performed, such as using pinned memory in OPM for the assembly part to avoid the implicit CPU-to-CPU memory transfers of the system matrix before the data is copied to the GPU, and overlapping the copy of the original matrix to the GPU while the Jacobi matrix is created on the CPU. 

\section*{Acknowledgments}
\label{sec:acknowledgments}

We gratefully acknowledge the support of EPIC – Energy Production Innovation Center, hosted by the University of Campinas (UNICAMP) and sponsored by FAPESP – São Paulo Research Foundation (2017/15736-3). We acknowledge the support of ANP (Brazil’s National Oil, Natural Gas and Biofuels Agency) through the R\&D levy regulation - Project reference number 24174-5. Special thanks are given to the "Centro de Estudos de Energia e Petróleo" (CEPETRO), School of Mechanical Engineering at UNICAMP and (other Institutes of UNICAMP or other University) for their support throughout this research.

\bibliography{sample}

\begin{thebibliography}{10}
\providecommand{\url}[1]{#1}
\csname url@samestyle\endcsname
\providecommand{\newblock}{\relax}
\providecommand{\bibinfo}[2]{#2}
\providecommand{\BIBentrySTDinterwordspacing}{\spaceskip=0pt\relax}
\providecommand{\BIBentryALTinterwordstretchfactor}{4}
\providecommand{\BIBentryALTinterwordspacing}{\spaceskip=\fontdimen2\font plus
\BIBentryALTinterwordstretchfactor\fontdimen3\font minus \fontdimen4\font\relax}
\providecommand{\BIBforeignlanguage}[2]{{%
\expandafter\ifx\csname l@#1\endcsname\relax
\typeout{** WARNING: IEEEtran.bst: No hyphenation pattern has been}%
\typeout{** loaded for the language `#1'. Using the pattern for}%
\typeout{** the default language instead.}%
\else
\language=\csname l@#1\endcsname
\fi
#2}}
\providecommand{\BIBdecl}{\relax}
\BIBdecl

\bibitem{OPM}
\BIBentryALTinterwordspacing
 [Online]. Available: \url{https://opm-project.org}
\BIBentrySTDinterwordspacing

\bibitem{dune}
\BIBentryALTinterwordspacing
M.~Blatt, A.~Burchardt, A.~Dedner, C.~Engwer, J.~Fahlke, B.~Flemisch, C.~Gersbacher, C.~Gr\"aser, F.~Gruber, C.~Gr\"uninger, D.~Kempf, R.~Kl{\"o}fkorn, T.~Malkmus, S.~Müthing, M.~Nolte, M.~Piatkowski, and O.~Sander, ``{The Distributed and Unified Numerics Environment, Version 2.4},'' \emph{Archive of Numerical Software}, vol.~4, no. 100, pp. 13--29, 2016. [Online]. Available: \url{http://dx.doi.org/10.11588/ans.2016.100.26526}
\BIBentrySTDinterwordspacing

\bibitem{ILU0-opt}
\BIBentryALTinterwordspacing
A.~Thune, X.~Cai, and A.~B. Rustad, ``On the impact of heterogeneity-aware mesh partitioning and non-contributing computation removal on parallel reservoir simulations,'' \emph{Journal of Mathematics in Industry}, 2021. [Online]. Available: \url{https://mathematicsinindustry.springeropen.com/articles/10.1186/s13362-021-00108-5}
\BIBentrySTDinterwordspacing

\bibitem{CPR}
\BIBentryALTinterwordspacing
K.~Wang, H.~Liu, J.~Luo, and Z.~Chen, ``Efficient cpr-type preconditioner and its adaptive strategies for large-scale parallel reservoir simulations,'' \emph{Journal of Computational and Applied Mathematics}, vol. 328, pp. 443--468, 2018. [Online]. Available: \url{https://www.sciencedirect.com/science/article/pii/S0377042717303734}
\BIBentrySTDinterwordspacing

\bibitem{gpu_weather}
J.~Michalakes and M.~Vachharajani, ``Gpu acceleration of numerical weather prediction,'' \emph{Parallel Processing Letters}, vol.~18, no.~04, pp. 531--548, 2008.

\bibitem{gpu_struct}
S.~Georgescu, P.~Chow, and H.~Okuda, ``Gpu acceleration for fem-based structural analysis,'' \emph{Archives of Computational Methods in Engineering}, vol.~20, no.~2, pp. 111--121, 2013.

\bibitem{gpu_ode}
L.~Murray, ``Gpu acceleration of runge-kutta integrators,'' \emph{IEEE Transactions on Parallel and Distributed Systems}, vol.~23, no.~1, pp. 94--101, 2012.

\bibitem{gpu_chip}
Y.~Lin, ``Gpu acceleration in vlsi back-end design: Overview and case studies,'' in \emph{2020 IEEE/ACM International Conference On Computer Aided Design (ICCAD)}, 2020, pp. 1--4.

\bibitem{gpu_tensorflow}
\BIBentryALTinterwordspacing
M.~Abadi, A.~Agarwal, P.~Barham, E.~Brevdo, Z.~Chen, C.~Citro, G.~S. Corrado, A.~Davis, J.~Dean, M.~Devin, S.~Ghemawat, I.~Goodfellow, A.~Harp, G.~Irving, M.~Isard, Y.~Jia, R.~Jozefowicz, L.~Kaiser, M.~Kudlur, J.~Levenberg, D.~Man\'{e}, R.~Monga, S.~Moore, D.~Murray, C.~Olah, M.~Schuster, J.~Shlens, B.~Steiner, I.~Sutskever, K.~Talwar, P.~Tucker, V.~Vanhoucke, V.~Vasudevan, F.~Vi\'{e}gas, O.~Vinyals, P.~Warden, M.~Wattenberg, M.~Wicke, Y.~Yu, and X.~Zheng, ``{TensorFlow}: Large-scale machine learning on heterogeneous systems,'' 2015, software available from tensorflow.org. [Online]. Available: \url{https://www.tensorflow.org/}
\BIBentrySTDinterwordspacing

\bibitem{flow_paper}
\BIBentryALTinterwordspacing
A.~F. Rasmussen, T.~H. Sandve, K.~Bao, A.~Lauser, J.~Hove, B.~Skaflestad, R.~Klöfkorn, M.~Blatt, A.~B. Rustad, O.~Sævareid, K.-A. Lie, and A.~Thune, ``The open porous media flow reservoir simulator,'' 2019. [Online]. Available: \url{https://arxiv.org/abs/1910.06059}
\BIBentrySTDinterwordspacing

\bibitem{reservoir_sim1}
J.~H. Abou-Kassem, S.~M. Farouq-Ali, and M.~R. Islam, \emph{Petroleum reservoir simulations}.\hskip 1em plus 0.5em minus 0.4em\relax Elsevier, 2013.

\bibitem{reservoir_sim2}
\BIBentryALTinterwordspacing
 [Online]. Available: \url{https://petrowiki.spe.org/Reservoir_simulation}
\BIBentrySTDinterwordspacing

\bibitem{blackoil_model}
\BIBentryALTinterwordspacing
J.~A. Trangenstein and J.~B. Bell, ``Mathematical structure of the black-oil model for petroleum reservoir simulation,'' \emph{SIAM Journal on Applied Mathematics}, vol.~49, pp. 749--783, 1989. [Online]. Available: \url{http://doi.org/10.2307/2101984}
\BIBentrySTDinterwordspacing

\bibitem{eclipse}
\BIBentryALTinterwordspacing
 [Online]. Available: \url{https://www.software.slb.com/products/eclipse}
\BIBentrySTDinterwordspacing

\bibitem{resinsight}
\BIBentryALTinterwordspacing
 [Online]. Available: \url{https://resinsight.org/}
\BIBentrySTDinterwordspacing

\bibitem{darcy_law1}
M.~Muskat and M.~W. Meres, ``The flow of heterogeneous fluids through porous media,'' \emph{Physics}, vol.~7, no.~9, pp. 346--363, 1936.

\bibitem{darcy_law2}
M.~Tek, ``Development of a generalized darcy equation,'' \emph{Journal of Petroleum Technology}, vol.~9, no.~06, pp. 45--47, 1957.

\bibitem{flow_manual}
\BIBentryALTinterwordspacing
 [Online]. Available: \url{https://opm-project.org/?page_id=955}
\BIBentrySTDinterwordspacing

\bibitem{dune-istl}
M.~Blatt and P.~Bastian, ``The iterative solver template library,'' in \emph{Applied Parallel Computing -- State of the Art in Scientific Computing}, B.~Kagstr\"om, E.~Elmroth, J.~Dongarra, and J.~Wasniewski, Eds.\hskip 1em plus 0.5em minus 0.4em\relax Berlin/Heidelberg: Springer, 2007, pp. 666--675.

\bibitem{boast1}
\BIBentryALTinterwordspacing
J.~R. Fanchi, K.~J. Harpole, and S.~W. Bujnowski, ``Boast: a three-dimensional, three-phase black oil applied simulation tool (version 1. 1). volume i. technical description and fortran code,'' 9 1982. [Online]. Available: \url{https://www.osti.gov/biblio/7069892}
\BIBentrySTDinterwordspacing

\bibitem{boast2}
\BIBentryALTinterwordspacing
 [Online]. Available: \url{https://netl.doe.gov/node/7530}
\BIBentrySTDinterwordspacing

\bibitem{mrst}
K.-A. Lie, \emph{An Introduction to Reservoir Simulation Using MATLAB/GNU Octave: User Guide for the MATLAB Reservoir Simulation Toolbox (MRST)}.\hskip 1em plus 0.5em minus 0.4em\relax Cambridge University Press, 2019.

\bibitem{intersect}
\BIBentryALTinterwordspacing
 [Online]. Available: \url{https://www.software.slb.com/products/intersect}
\BIBentrySTDinterwordspacing

\bibitem{echelon}
\BIBentryALTinterwordspacing
 [Online]. Available: \url{http://stoneridgetechnology.com/echelon/}
\BIBentrySTDinterwordspacing

\bibitem{terapowers}
\BIBentryALTinterwordspacing
 [Online]. Available: \url{https://www.aramco.com/en/creating-value/technology-development/in-house-developed-technologies/terapowers}
\BIBentrySTDinterwordspacing

\bibitem{shaheen}
\BIBentryALTinterwordspacing
 [Online]. Available: \url{https://www.hpc.kaust.edu.sa/content/shaheen-ii}
\BIBentrySTDinterwordspacing

\bibitem{geosx}
\BIBentryALTinterwordspacing
 [Online]. Available: \url{http://www.geosx.org/}
\BIBentrySTDinterwordspacing

\bibitem{tnavigator}
\BIBentryALTinterwordspacing
 [Online]. Available: \url{http://tnavigator.com/}
\BIBentrySTDinterwordspacing

\bibitem{pflotran}
P.~C. Lichtner, G.~E. Hammond, C.~Lu, S.~Karra, G.~Bisht, B.~Andre, R.~T. Mills, J.~Kumar, and J.~M. Frederick, ``{PFLOTRAN} {W}eb page,'' 2020, http://www.pflotran.org.

\bibitem{cusparse}
\BIBentryALTinterwordspacing
 [Online]. Available: \url{https://docs.nvidia.com/cuda/cusparse/index.html}
\BIBentrySTDinterwordspacing

\bibitem{magma}
\BIBentryALTinterwordspacing
 [Online]. Available: \url{https://icl.utk.edu/magma/index.html}
\BIBentrySTDinterwordspacing

\bibitem{magma_paper}
H.~Anzt, W.~Sawyer, S.~Tomov, P.~Luszczek, I.~Yamazaki, and J.~Dongarra, ``Optimizing krylov subspace solvers on graphics processing units,'' in \emph{Fourth International Workshop on Accelerators and Hybrid Exascale Systems (AsHES), IPDPS 2014}, IEEE.\hskip 1em plus 0.5em minus 0.4em\relax Phoenix, AZ: IEEE, 05-2014 2014.

\bibitem{viennacl}
\BIBentryALTinterwordspacing
 [Online]. Available: \url{http://viennacl.sourceforge.net/}
\BIBentrySTDinterwordspacing

\bibitem{ginkgo-toms-2022}
\BIBentryALTinterwordspacing
H.~Anzt, T.~Cojean, G.~Flegar, F.~Göbel, T.~Grützmacher, P.~Nayak, T.~Ribizel, Y.~M. Tsai, and E.~S. Quintana-Ortí, ``{Ginkgo: A Modern Linear Operator Algebra Framework for High Performance Computing},'' \emph{ACM Transactions on Mathematical Software}, vol.~48, no.~1, pp. 2:1--2:33, Feb. 2022. [Online]. Available: \url{https://doi.org/10.1145/3480935}
\BIBentrySTDinterwordspacing

\bibitem{amgcl1}
\BIBentryALTinterwordspacing
D.~Demidov, ``Amgcl: An efficient, flexible, and extensible algebraic multigrid implementation,'' \emph{Lobachevskii Journal of Mathematics}, vol.~40, no.~5, pp. 535--546, May 2019. [Online]. Available: \url{https://doi.org/10.1134/S1995080219050056}
\BIBentrySTDinterwordspacing

\bibitem{rocalution}
\BIBentryALTinterwordspacing
 [Online]. Available: \url{https://github.com/ROCmSoftwarePlatform/rocALUTION}
\BIBentrySTDinterwordspacing

\bibitem{rocsparse}
\BIBentryALTinterwordspacing
 [Online]. Available: \url{https://github.com/ROCmSoftwarePlatform/rocSPARSE}
\BIBentrySTDinterwordspacing

\bibitem{hip}
\BIBentryALTinterwordspacing
 [Online]. Available: \url{https://github.com/ROCm-Developer-Tools/HIP}
\BIBentrySTDinterwordspacing

\bibitem{rocm}
\BIBentryALTinterwordspacing
 [Online]. Available: \url{https://github.com/RadeonOpenCompute/ROCm}
\BIBentrySTDinterwordspacing

\bibitem{chow_patel_decomp}
E.~Chow and A.~Patel, ``Fine-grained parallel incomplete lu factorization,'' \emph{SIAM journal on Scientific Computing}, vol.~37, no.~2, pp. C169--C193, 2015.

\bibitem{chow_patel_apply}
H.~Anzt, E.~Chow, and J.~Dongarra, ``Iterative sparse triangular solves for preconditioning,'' in \emph{Euro-Par 2015: Parallel Processing}, J.~L. Tr{\"a}ff, S.~Hunold, and F.~Versaci, Eds.\hskip 1em plus 0.5em minus 0.4em\relax Berlin, Heidelberg: Springer Berlin Heidelberg, 2015, pp. 650--661.

\bibitem{blocked_spmv}
R.~Eberhardt and M.~Hoemmen, ``Optimization of block sparse matrix-vector multiplication on shared-memory parallel architectures,'' in \emph{2016 IEEE International Parallel and Distributed Processing Symposium Workshops (IPDPSW)}.\hskip 1em plus 0.5em minus 0.4em\relax IEEE, 2016, pp. 663--672.

\bibitem{Thune_2021}
\BIBentryALTinterwordspacing
A.~Thune, X.~Cai, and A.~B. Rustad, ``On the impact of heterogeneity-aware mesh partitioning and non-contributing computation removal on parallel reservoir simulations,'' \emph{Journal of Mathematics in Industry}, vol.~11, no.~1, jun 2021. [Online]. Available: \url{https://doi.org/10.1186/s13362-021-00108-5}
\BIBentrySTDinterwordspacing

\bibitem{andreas}
\BIBentryALTinterwordspacing
 [Online]. Available: \url{https://github.com/andrthu}
\BIBentrySTDinterwordspacing

\bibitem{umfpack}
\BIBentryALTinterwordspacing
T.~A. Davis, ``Algorithm 832: Umfpack v4.3---an unsymmetric-pattern multifrontal method,'' \emph{ACM Trans. Math. Softw.}, vol.~30, no.~2, p. 196–199, jun 2004. [Online]. Available: \url{https://doi.org/10.1145/992200.992206}
\BIBentrySTDinterwordspacing

\bibitem{amgcl2}
\BIBentryALTinterwordspacing
D.~Demidov, ``Amgcl -- a c++ library for efficient solution of large sparse linear systems,'' \emph{Software Impacts}, vol.~6, p. 100037, 2020. [Online]. Available: \url{https://doi.org/10.1016/j.simpa.2020.100037}
\BIBentrySTDinterwordspacing

\bibitem{vexcl}
\BIBentryALTinterwordspacing
 [Online]. Available: \url{https://github.com/ddemidov/vexcl}
\BIBentrySTDinterwordspacing

\bibitem{eigen}
\BIBentryALTinterwordspacing
 [Online]. Available: \url{http://eigen.tuxfamily.org/}
\BIBentrySTDinterwordspacing

\bibitem{blaze}
\BIBentryALTinterwordspacing
 [Online]. Available: \url{https://bitbucket.org/blaze-lib/blaze}
\BIBentrySTDinterwordspacing

\bibitem{cublas}
\BIBentryALTinterwordspacing
 [Online]. Available: \url{https://docs.nvidia.com/cuda/cublas/index.html}
\BIBentrySTDinterwordspacing

\bibitem{norne}
\BIBentryALTinterwordspacing
 [Online]. Available: \url{https://opm-project.org/?page_id=559}
\BIBentrySTDinterwordspacing

\bibitem{simula1}
\BIBentryALTinterwordspacing
 [Online]. Available: \url{https://www.simula.no/}
\BIBentrySTDinterwordspacing

\bibitem{simula2}
\BIBentryALTinterwordspacing
 [Online]. Available: \url{https://www.ex3.simula.no/}
\BIBentrySTDinterwordspacing

\bibitem{simula3}
\BIBentryALTinterwordspacing
 [Online]. Available: \url{https://www.simula.no/news/improving-norwegian-research-infrastructure-experimental-infrastructure-exploration-exascale}
\BIBentrySTDinterwordspacing

\bibitem{simula4}
\BIBentryALTinterwordspacing
 [Online]. Available: \url{http://wiki.ex3.simula.no/doku.php}
\BIBentrySTDinterwordspacing

\bibitem{own_paper}
\BIBentryALTinterwordspacing
T.~Hogervorst, R.~Nane, G.~Marchiori, T.~D. Qiu, M.~Blatt, and A.~B. Rustad, ``Hardware acceleration of high-performance computational flow dynamics using high-bandwidth memory-enabled field-programmable gate arrays,'' \emph{ACM Trans. Reconfigurable Technol. Syst.}, vol.~15, no.~2, dec 2021. [Online]. Available: \url{https://doi.org/10.1145/3476229}
\BIBentrySTDinterwordspacing

\bibitem{opmrepo}
\BIBentryALTinterwordspacing
OPM, ``Open porous media project,'' 2020. [Online]. Available: \url{https://github.com/OPM}
\BIBentrySTDinterwordspacing

\end{thebibliography}

\end{document}